\documentclass[runningheads]{llncs}
\usepackage[T1]{fontenc}
\usepackage{graphicx}
\usepackage{wrapfig}
\usepackage{xcolor}
\usepackage{makecell}
\usepackage{multirow}
\usepackage{hyperref}

\begin{document}
\title{Modular AI-Powered Interviewer with Dynamic Question Generation and Expertise Profiling}
\titlerunning{Modular AI-Powered Interviewer}
%
\author{Aisvarya Adeseye\inst{1}\orcidID{0009-0003-2401-3076} \and
Jouni Isoaho\inst{1}\orcidID{0000-0002-5789-3992} \and
Seppo Virtanen\inst{1}\orcidID{0000-0002-9487-3018}
\and
Mohammad Tahir\inst{1}\orcidID{0000-0002-6273-4603}}
\authorrunning{A. Adeseye et al.}
%
\institute{Department of Computing, University of Turku, Turku, Finland \\
\email{\{aisvarya.a.adeseye, jouni.isoaho, seppo.virtanen, tahir.mohammad\}@utu.fi}
}

\maketitle              
\begin{abstract}

Automated interviewers and chatbots are common in research, recruitment, customer service, and education. Many existing systems use fixed question lists, strict rules, and limited personalization, leading to repeated conversations that cause low engagement. Therefore, these tools are not effective for complex qualitative research, which requires flexibility, context awareness, and ethical sensitivity. Consequently, there is a need for a more adaptive and context-aware interviewing system. To address this, an AI-powered interviewer that dynamically generates questions that are contextually appropriate and expertise aligned is presented in this study. The interviewer is built on a locally hosted large language model (LLM) that generates coherent dialogue while preserving data privacy. The interviewer profiles the participants' expertise in real time to generate knowledge-appropriate questions, well-articulated responses, and smooth transition messages similar to human-like interviews. To implement these functionalities, a modular prompt engineering pipeline was designed to ensure that the interview conversation remains scalable, adaptive, and semantically rich. To evaluate the AI-powered interviewer, it was tested with various participants, and it achieved high satisfaction (mean 4.45) and engagement (mean 4.33). The proposed interviewer is a scalable, privacy-conscious solution that advances AI-assisted qualitative data collection.

\end{abstract}
\section{Introduction}

Artificial Intelligence (AI) plays a vital role in improving human–machine interaction in domains such as customer service, education, recruitment, and research. 

Chatbots \cite{han2021} like like Replika \cite{folstad2021}, Mitsuku \cite{auer2024}, and automated interviewers \cite{ajunwa2021} are tools that help make interaction consistent, reducing human workload. Chatbots use predefined rules or intent recognition to handle simple and routine questions. They have limited flexibility, hence, they are not adaptive to the expertise level of the user.

Contrariwise, automated interviewers such as HireVue \cite{drage2022} and Tengai \cite{bolin2024} are early efforts to achieve automated interviews. They are designed to handle structured or semi-structured interactions in recruitment, research, or evaluation. However, they mostly rely on traditional Natural Language Processing (NLP) techniques, which limit their capacity to generate adaptive and contextually relevant follow-up questions.

Although chatbots and automated interviewers are broadly adopted, they generally suffer from poor personalization, inflexible scripts that make dialogues repetitive. In the last few years, large language models  (LLMs) like GPT, LLaMA, Gemini, and DeepSeek have emerged, enabling dynamic question generation, adaptive follow-ups in conversational dialogue without the need for hard-coded templates. 

However, LLM models have their own challenges, such as producing irrelevant information, hallucinations, and redundant responses. To mitigate these, effective prompt designs are needed to ensure that the model produces clear, accurate, and relevant responses. Furthermore, a structured prompt design and appropriate performance evaluation are two important aspects to ensure that LLM-based interviews stay context-aware and reliable. 

Consequently, this study aims to design and analyze the performance of an adaptive conversational AI-powered interviewer that generates contextually relevant and non-repetitive interview questions. Additionally, it evaluates how well the system maintains coherent and user-appropriate dialogue flow with participants of diverse knowledge, experience, and expertise levels.

Hence, this study focuses on two main research questions (RQs):

\begin{itemize}
\item \textbf{RQ1-}How effectively does an AI-powered interviewer generate a coherent flow of contextually appropriate, expertise-aligned questions across interview sessions through prompt-driven logic?

\item \textbf{RQ2-}What are the performance, adaptability, and overall user experience of an AI-powered interviewer?
\end{itemize}

This research provides two key contributions which are (1) a modular, prompt-driven AI-powered interviewer that dynamically adjusts question complexity in real time based on the user’s expertise level, and (2) a structured framework for generating contextually relevant, non-redundant, and semantically coherent interview dialogues using locally hosted LLMs.

\section{Background Study}

\textbf{Existing Approaches in Chatbots and Automated Interviewers}

Chatbots and automated interviewers are two different tools widely used for conversational interaction between human and machine \cite{han2021}. Chatbots are designed to do routine tasks, and are often used for answering frequently asked questions, scheduling appointments, and giving scripted customer support \cite{nirala2022}. Examples such as Replika \cite{folstad2021}  and Mitsuku \cite{auer2024} use rule-based logic or intent classifiers to respond. Their ability to personalize conversations or to ask nuanced follow-up questions is limited because of their static designs.

However, automated interviewers are more suited and generally used for structured interactions, such as candidate screening, academic interviews, and qualitative research \cite{langer2020}. Systems such as HireVue \cite{drage2022} and Tengai \cite{bolin2024} attempt to make interviews like real conversations. But they have limitations such as asking repeated questions, not being able to understand conversations well, and a lack of adaptability to different scenarios due to rigid question structure or limited NLP capabilities. The AI-powered interviewer proposed in this study addresses these limitations by integrating local LLMs to dynamically generate context-aware and expertise-aligned interviews.

\textbf{Intelligent Interview Technologies}

NLP is a widely used technique in the development of conversational agents \cite{sharma2024}. Basic interactions in earlier systems used template-based generation, part-of-speech tagging, and entity recognition \cite{nasar2021}. However, due to the evolution of NLP techniques, methods such as topic modeling, semantic similarity scoring, and sequence-to-sequence models were introduced to improve the conversational flow and thematic relevance \cite{Pandey2023}. Also, some interview bots incorporate domain-specific ontologies or knowledge graphs to ensure question consistency within technical fields \cite{abusalih2021}.

However, these systems have limited scalability because they heavily depend on manually designed dialogue flows, which limit their responsiveness to unexpected user inputs. The popularity of LLMs has improved conversational dialogue. Models like GPT-4, LLaMA, and Gemini can generate human-like but coherent conversation that adapts to the user’s domain, tone, or expertise level without needing predefined templates \cite{veeramachaneni2025}.

Despite these benefits, cloud-based commercial LLMs cannot be used to process sensitive information due to privacy and regulatory compliance. Examples include ChatGPT and Gemini. Consequently, there is growing interest in deploying LLMs locally in institutional or personal infrastructure, which guarantees data privacy, the blocking of third-party access, and maintains ethical considerations in data-sensitive research and operations.

\textbf{The Role of Prompt Engineering in Local LLM}

Local LLMs are highly sensitive to prompt structure unlike commercial models that use a wide range of reinforcement training and server-side optimization to improve output accuracy and quality. Without these techniques, prompt engineering is critical to extract reliable outputs from local models. 

When using local language models, prompt engineering is very important because they do not have the extensive fallback mechanisms as commercial models \cite{liu2025}. To get effective results, local models need clear and well-structured prompts. If the prompts are poorly designed, the model may produce confusing or irrelevant outputs. In worst cases, it may lead to hallucination, contradictions, or a break in the conversational flow.

There are two main types of prompts that control how an LLM behaves \cite{adeseye2025ichms}. System prompts define the model behavior during the entire session \cite{giray2023}. For example, a system prompt might instruct the model to provide responses according to the user's preferences. User prompts are specific inputs given during interaction, such as responses from the user or situational data \cite{jishan2024}. They assist the model in deciding what to ask next (follow-up questions) or summarize what happened earlier.

The main difference between system and user prompts is about what they do: system prompt helps the model understand its goals and identity. However, user prompts help the model answer questions in real-time. When they work together, they create outputs with minimal repetition, dynamic, highly relevant, and sensitive to user input. Therefore, to design an effective AI-powered interviewer, prompt engineering is essential

\section{AI-Powered Interviewer Architecture}

\begin{figure}
    \centering
    \includegraphics[width=0.8\linewidth]{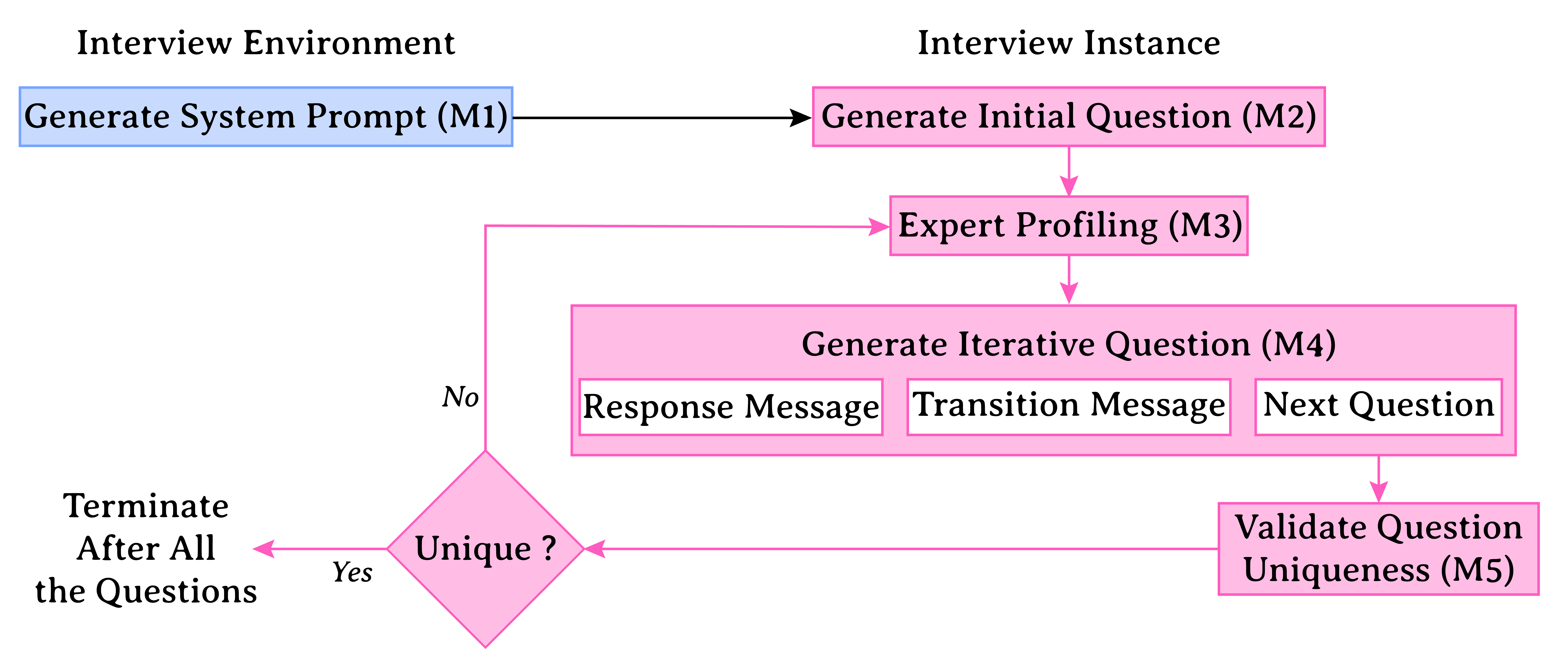}
    \caption{Modular Pipeline Architecture of AI-Powered Interviewer }
    \label{fig:InterviewArchitecture}
\end{figure}

The presented AI-powered Interviewer is a locally deployed system that assists with interviews. It is designed to understand and communicate with humans. The system is made up of small, reusable parts called modules. Each module does a specific job and passes its output to the next module. This modular architecture makes the interviewer flexible, reusable, and transparent. The modules are independent and can be added or removed without affecting the entire system. This architecture makes it easy to adapt to different cultures or languages, change the level of difficulty for participants, and add or remove features easily. The five main modules and their interaction are illustrated in Figure \ref{fig:InterviewArchitecture}.

This modular architecture design has several advantages from an implementation perspective. It assists in separating different tasks, making it easier to optimize, fix bugs, and modify each part for different purposes independently, for example, different interview topics or ethical rules. The architecture is also transparent, making it easy to inspect and audit the outputs from each module.

\subsection{Module 1 (M1): Generate System Prompt}

\begin{figure}
    \centering
    \includegraphics[width=0.8\linewidth]{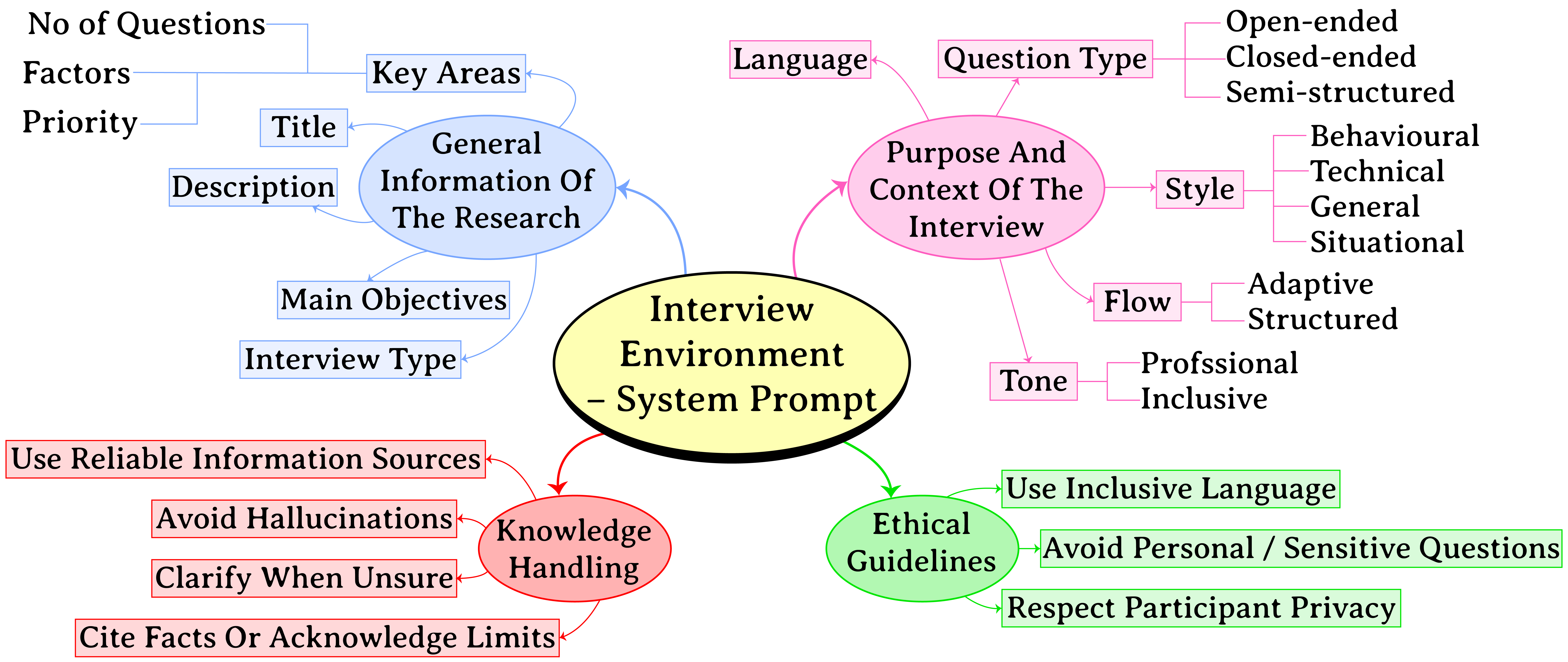}
    \caption{System Prompt Outline in Generating the System Prompt Module}
    \label{fig:module1-SP}
\end{figure}

\begin{figure}
    \centering
    \includegraphics[width=0.7\linewidth]{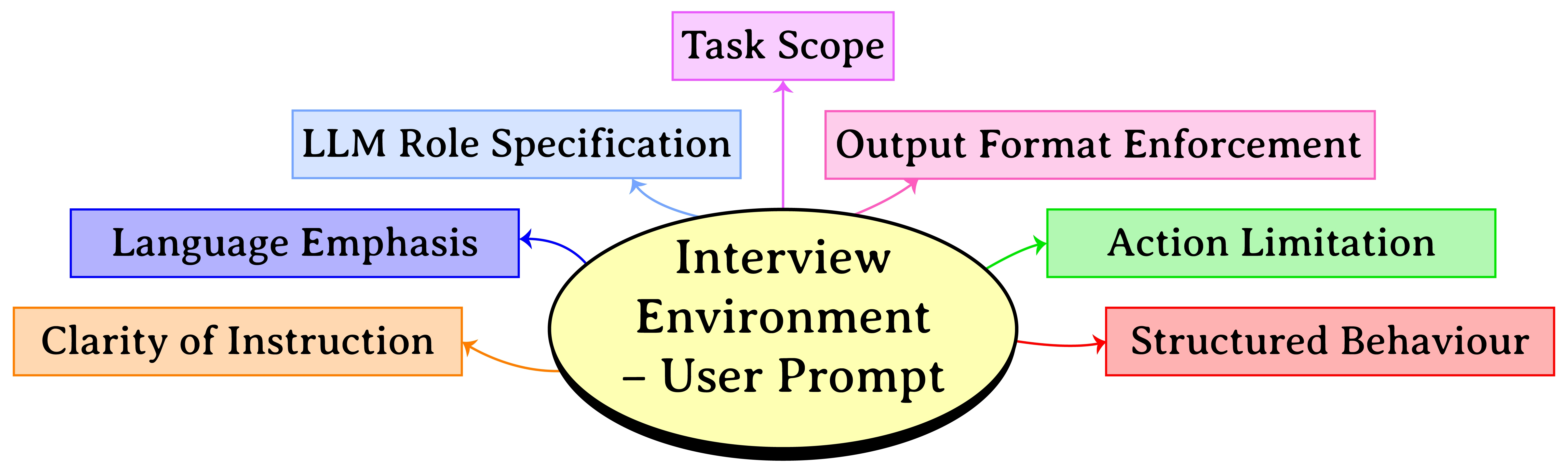}
    \caption{User Prompt Outline in Generating the System Prompt Module}
    \label{fig:module1-UP}
\end{figure}

This module sets up the interview environment by creating a structured system prompt that tells the local LLM how to behave throughout the entire interview session. It receives information provided in figure \ref{fig:module1-SP} to generate a single reusable JSON prompt. This is the only time the system prompt is created and stored, so the interviewer provides consistent, ethical, and context-aware responses during the entire interaction. Figure \ref{fig:module1-SP} depicts the structure of the system prompt, while Figure \ref{fig:module1-UP} contains the structure of the user prompt used in generating the system prompt module.

\subsection{Module 2 (M2): Generate Initial Question}

\begin{figure}
    \centering
    \includegraphics[width=0.7\linewidth]{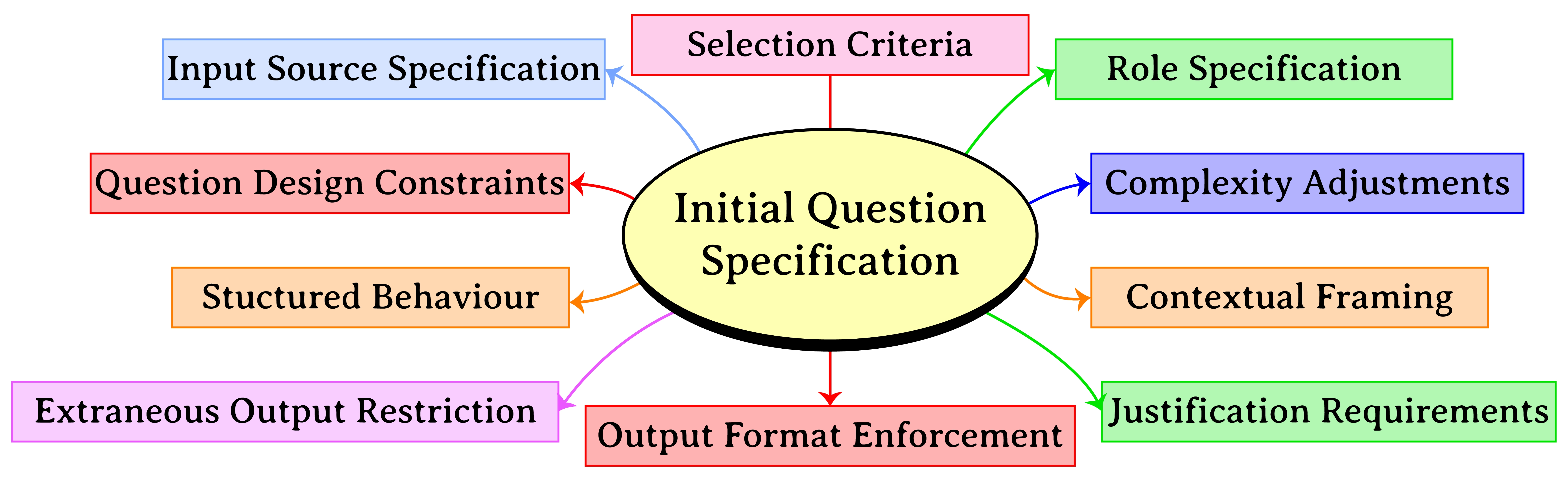}
    \caption{User Prompt Outline for Initial Question Prompt}
    \label{fig:module2-UP}
\end{figure}

The LLM begins the interview by selecting a high-priority key research area and generating a simple or less complex question that is easy for people to understand. This makes the interview accessible, encouraging early engagement. As seen in Figure \ref{fig:module2-UP}, this process is guided by these key elements:  the input source and role specification, complexity adjustments, contextual framing, justification requirements, and structured behavior. It also states "what format the answer should be" and "what not to say during the interview". These specifications ensure that the questions are crystal clear, purposeful, and aligned with the objective of the research.

\subsection{Module 3 (M3): Expertise Profiling}

\begin{figure}
    \centering
    \includegraphics[width=0.7\linewidth]{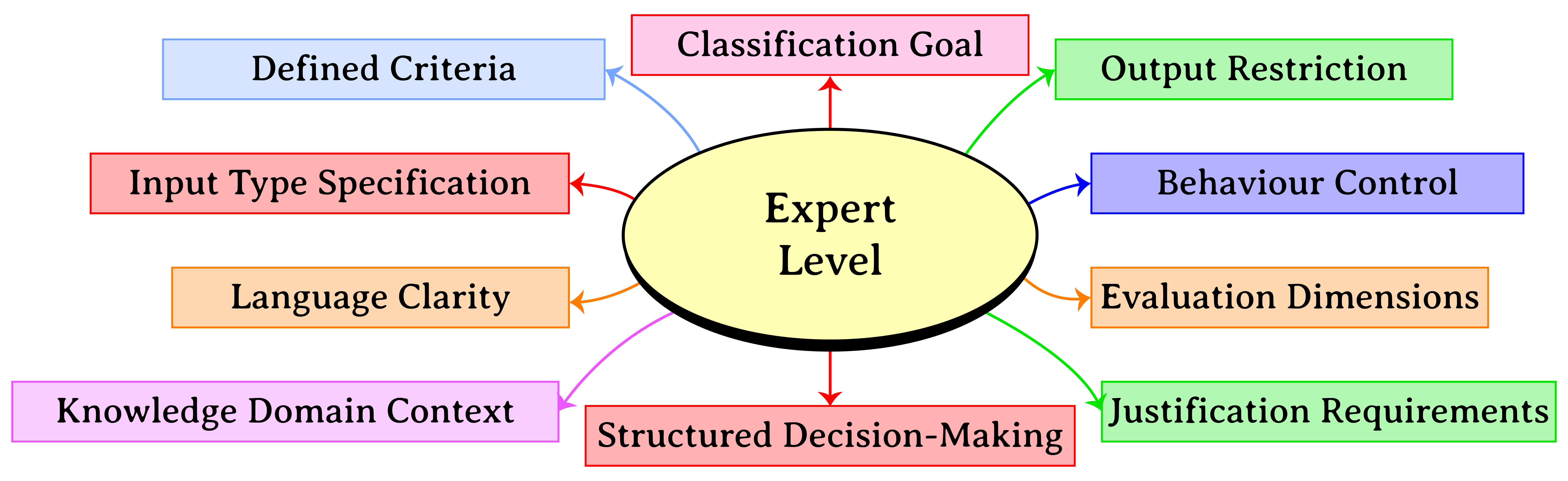}
    \caption{User Prompt Outline for Expert Level}
    \label{fig:module3-UP}
\end{figure}

This module profiles participants' expertise level via previous conversations. The LLM evaluates technical terminology, insight depth, and academic relevance with a four-level rubric, which includes \textit{Novice}, \textit{Basic Knowledge}, \textit{Advanced Knowledge}, and \textit{Expert}. As seen in Figure \ref{fig:module3-UP}, this classification depends on the following: input type specification, evaluation criteria, behavioral control, domain knowledge, and justification requirements. This module also helps Module 4 to create questions that adapt to the participant's expert level, ensuring that the interview remains challenging, interesting, and relevant so that it yields high-quality conversational data.

\subsection{Module 4 (M4): Generate Iterative Questions}

\begin{figure}
    \centering
    \includegraphics[width=0.7\linewidth]{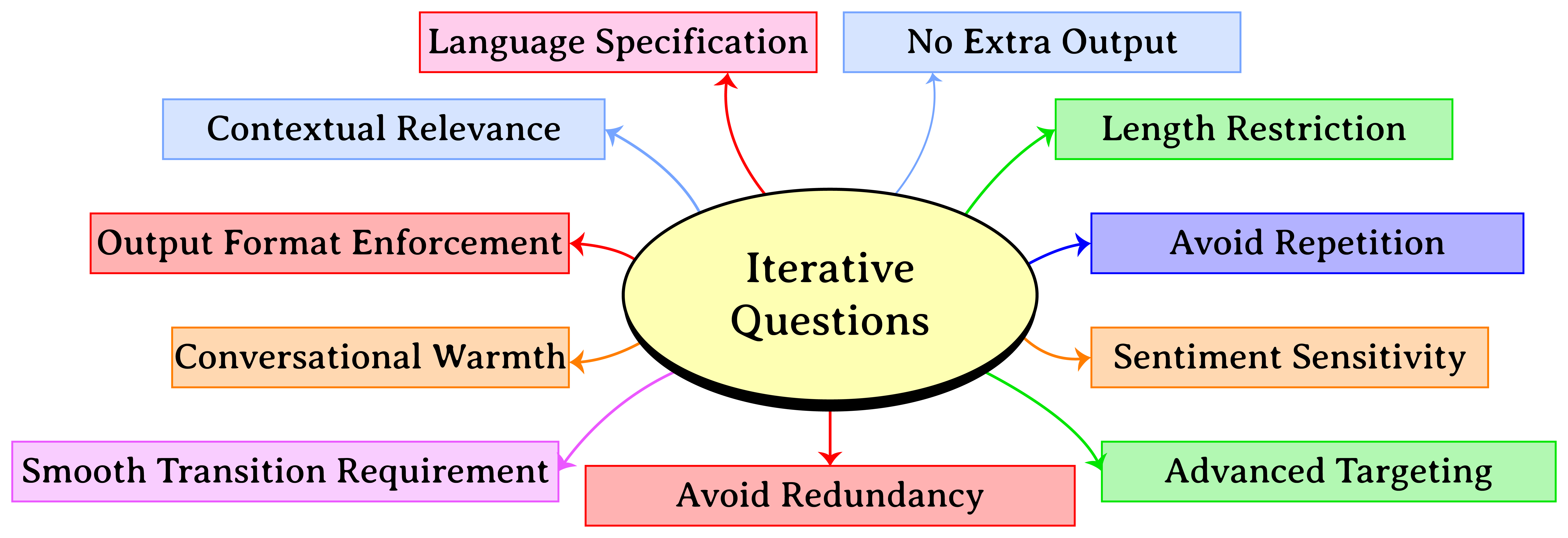}
    \caption{User Prompt Outline for Generate Iterative Questions}
    \label{fig:Module4-UP}
\end{figure}

This module creates follow-up questions that are adaptive, coherent, and conversational based on the participant's previous response and expertise level. It achieves this via three main structures: first, generating a brief response (under 10 words) that agrees or reflects on the participant's answer, second, write a smooth transition message to maintain the conversation flow, and lastly, formulate a context-aware, expertise-aligned, open-ended follow-up question. Additionally, it also includes a short justification (up to 25 words) on why a specific question was created. Figure \ref{fig:Module4-UP} shows the key constraints used by the LLM model to formulate the user prompt. This structure for the user prompt ensures that each question deepens the conversation and matches the participant's tone, conversation history, and topic continuity.

\subsection{Module 5 (M5): Validate Question Uniqueness}

\begin{figure}
    \centering
    \includegraphics[width=0.7\linewidth]{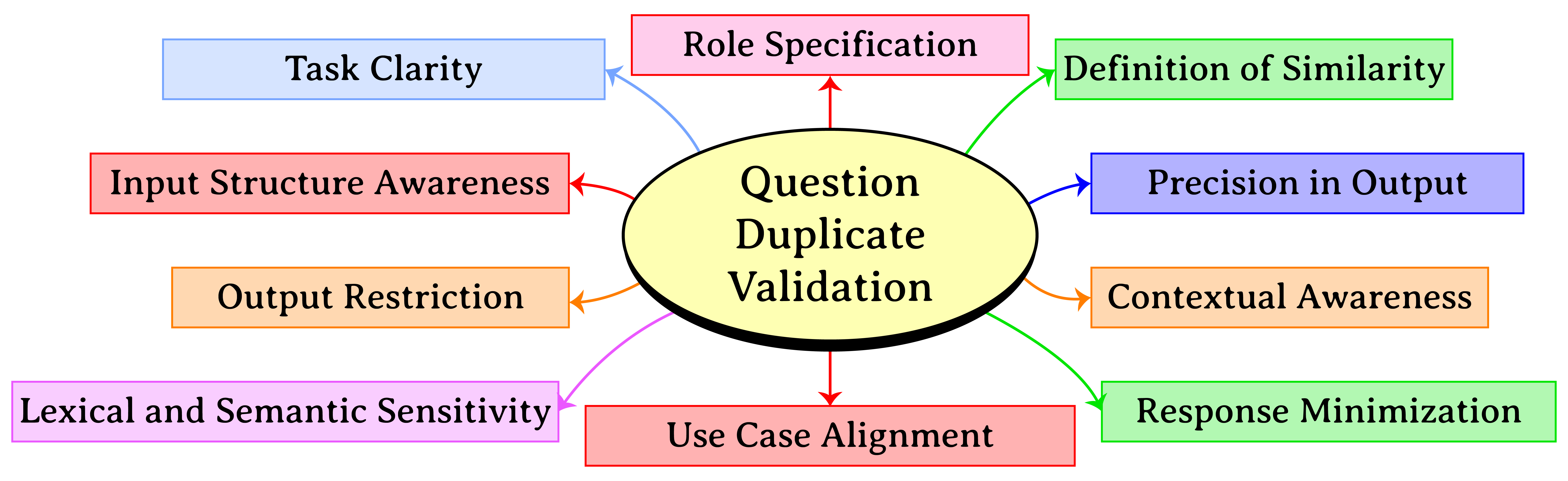}
    \caption{User Prompt Outline for Validate Question Uniqueness}
    \label{fig:Module5-UP}
\end{figure}

This module ensures that each new question is unique to avoid semantic duplication. The module checks the new question against previous ones to see if they mean the same thing (conceptual overlaps), even if they use different words, phrases, or synonymous substitutions. For example, if the model asks "What are LLMs?" and then the model again asks "What do you understand by Large Language Models?", the module would recognize that it is asking the same thing, even though the questions are phrased differently. Figure \ref{fig:Module5-UP} contains the key principles that need to be considered when designing the user prompt to implement this module. By incorporating these principles, the module ensures thematic progression, minimizes redundancy, and reduces participant fatigue. Also, this module helps collect reliable and accurate data.

\section{Technical Implementation}

A case study used to test the modular pipeline architecture of the AI-powered interviewer was about \textit{"how employees interact with LLMs in the workplace context"}. The study focused on employee awareness, skills, usage, and privacy concerns. The AI interviewer generates questions based on the research area name and priority level (High, Medium, Low), so the interview stays structured and relevant. Table \ref{tab:llm_research_priorities} contains the main research areas, their priority levels, and the number of questions for each research area for the case study.

\begin{table}
\caption{Priority Levels for Different Research Areas in LLM Awareness and Adoption.}\label{tab:llm_research_priorities}
\begin{tabular}{|l|l|l|}
\hline
\textbf{Research Area} & \textbf{Priority} & \textbf{No of Questions}\\
\hline
Awareness and knowledge of LLMs among employees & High & 4 \\
Application of LLMs in the Organization & Medium & 3\\
Skill levels and training in using LLMs & High & 3\\
Concerns related to data privacy and security in LLM use & Medium & 4\\
Organizational guidelines for LLM use and adoption & Low & 2\\
\hline
\end{tabular}
\end{table}

\begin{figure}
    \centering
    \includegraphics[width=1\linewidth]{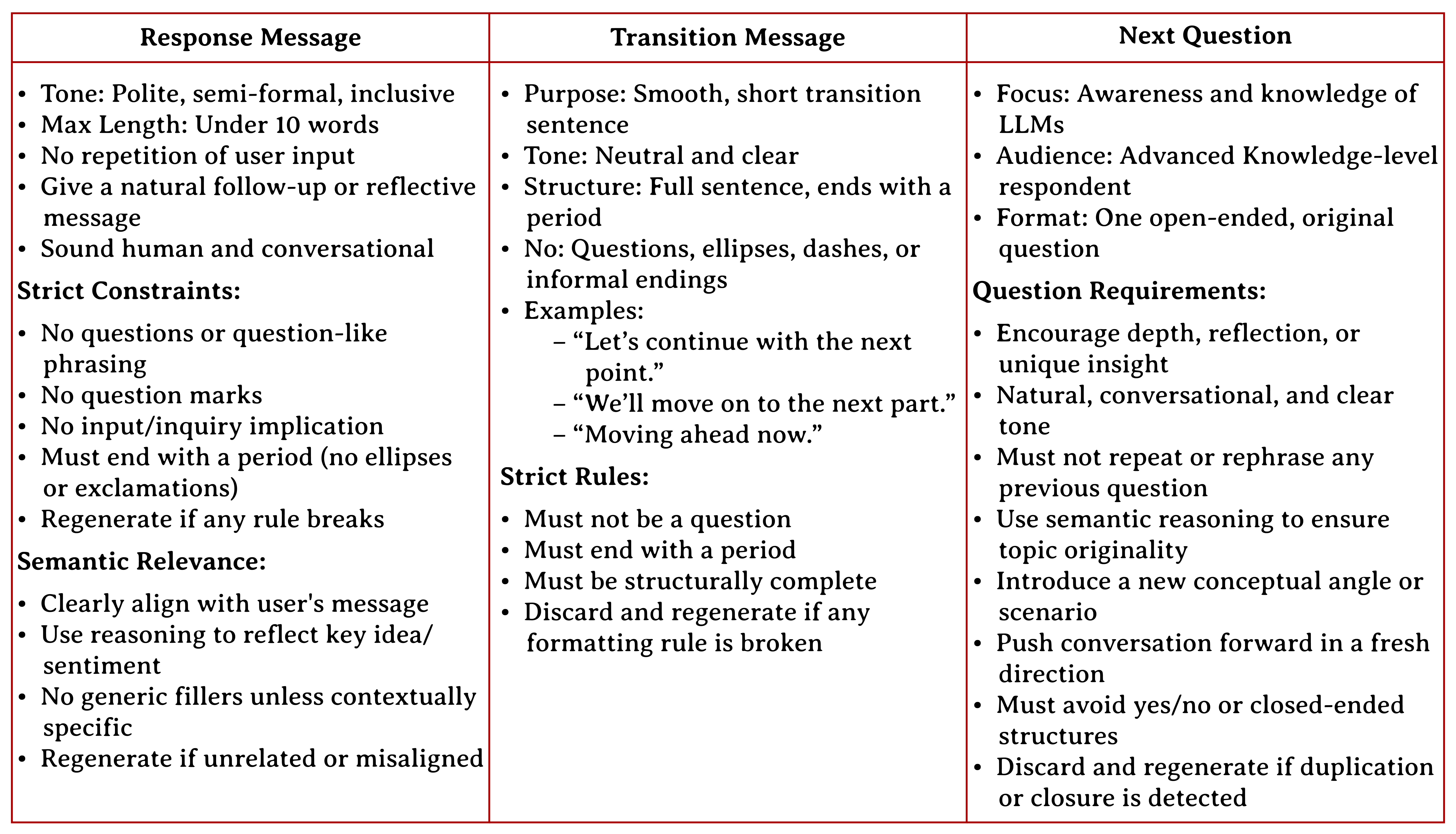}
    \caption{User Prompt Implementation Skeleton for Generative Iterative Questions (M4)}
    \label{fig:PromptDesign-IterativeQuestion}
\end{figure}

\subsection{Implementation}

LLaMA v3.2 (3B parameters) was used to implement the proposed modular pipeline architecture of the AI-powered interviewer. The prompts used to design all the modules in the architecture can be accessed through the following link: \href{https://github.com/aisvarya-adeseye/AI-Powered-Interviewer}{AI-Powered Interviewer GitHub Repository}. The simplified structure of the user prompt used in generating iterative questions (M4) is shown in Figure \ref{fig:PromptDesign-IterativeQuestion}.

\subsection{Testing and Results}

\begin{figure}
    \centering
    \includegraphics[width=0.6\linewidth]{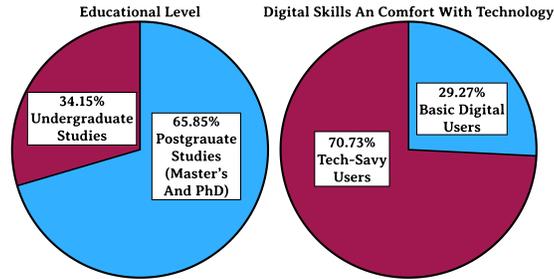}
    \caption{Demographic Information about the Pilot Testers}
    \label{fig:DEMOGRAPHICS}
\end{figure}

In the pilot study, a group of early users was selected to test the implemented AI-powered interviewer. Their feedback was important for refining the operations and functionality of the interviewer. Feedback received from 41 testers helped understand the strengths and potential areas for improvement. The testers were from diverse demographic, educational, and technical background (see Figure \ref{fig:DEMOGRAPHICS}). This attribute captured a wide range of real-world interactions, making the feedback robust and representative.

\subsubsection{Expertise Level}

The AI Interviewer demonstrated effective capability in dynamically identifying participants expertise level after each answer. Each interview started with a low-complexity, novice-friendly question. For example,

\begin{quote}
\scriptsize\textcolor{blue}{\textit{“Can you describe a situation where you've heard of or used a Large Language Model (LLM) before?”}}
\end{quote}

Based on the participant's response, the system assigned one of four expertise levels: Novice, Basic Knowledge, Advanced Knowledge, or Expert. Some participants started as Novices but moved to advanced knowledge as they showed increased knowledge and familiarity with the topic. For example, one participant began by talking about simple uses like summarizing reports, but later discussed the differences between local and commercial LLMs and the need for training within their organization for responsible use of LLMs. This shows that the system has the ability to recognize and adapt to how participants' expertise fluctuate during conversations.

\subsubsection{Question Complexity by Expertise Level}

The AI Interviewer adjusted the difficulty of the questions according to the expertise level of the participant. to keep the participant engaged and interested in the conversation,  allowing for a more in-depth and meaningful insight.

For novice-level participants, the system uses simple questions that are easy to understand and answer, for example:

\begin{quote}
\scriptsize\textcolor{blue}{\textit{“What do you think are some benefits or drawbacks of using AI tools in your daily work?”}}
\end{quote}

As the level of expertise increased, the question became relatively harder in terms of terminology and in-depth knowledge. Advanced knowledge users received questions like: 


\begin{quote}
\scriptsize\textcolor{blue}{\textit{“How do you think your organization should train employees to adapt to the presence of LLMs in daily operations?”}}
\end{quote}

At the Expert level, the system asked scenario-driven, ethics-oriented, or governance-related questions that require profound understanding of the subject matter:

\begin{quote}
\scriptsize\textcolor{blue}{\textit{“Can you describe a situation where a public LLM could pose a data security risk within your organization, and how you would mitigate it?”}}
\end{quote}

\subsubsection{Question Uniqueness}

The AI-powered Interviewer checks if newly generated questions are different from the ones that were already asked in the interview session. The interviewer ensures that no two follow-ups were semantically redundant. The follow-up questions were tailored based on the participants' background and experiences shared from prior responses. This ensures that the conversation stayed interesting, engaging, and contextually relevant.

\subsubsection{Progressive Adaptability}

The AI-powered Interviewer is effective at adapting to what participants said (progressive adaptation) by changing the way it asks questions, not just by adjusting the difficulty or complexity of the questions: the interview questions have a clear direction. It starts with general information and ends with deep thoughts on areas that include: responsible AI use, how organizations work, and ethical ways to use AI, capturing diverse participants' opinions and experiences. For example, at the beginning of the interview, one participant said this:

\begin{quote}
\scriptsize\textcolor{blue}{\textit{“I’ve heard of ChatGPT and used it to write summaries.”}}
\end{quote}

But mid-interview, the conversation was as follows:

\begin{quote}
\scriptsize\textcolor{blue}{\textit{“Organizations should train employees on local versus publicly available LLMs, and create internal guidelines on how data should be handled to mitigate ethical breaches.”}}
\end{quote}

This indicates that the interviewer understands conversation, adjusting questions complexity as the conversation proceeds. Also, the conversation flows freely without following any strict plan. 

\subsubsection{Response Message Accuracy}

This is an important but often overlooked part of the conversation, the \textit{response message}. It is a short and affirmative statement that confirms that the AI interviewer understood the response provided by the participant to the previous question in the conversation. The system consistently answered with semantically accurate and tone-appropriate responses. Examples include:

\begin{itemize}
  \item \scriptsize\textbf{Participant:} \scriptsize\textcolor{blue}{\textit{“I think AI tools may one day take over jobs.”}} \\
  \textbf{System Response:} \scriptsize\textcolor{blue}{\textit{“That’s a valid concern about job displacement.”}}
  
  \item \textbf{Participant:} \scriptsize\textcolor{blue}{\textit{“We use LLMs to optimize code in healthcare applications.”}} \\
  \textbf{System Response:} \scriptsize\textcolor{blue}{\textit{“That’s a great use-case for efficiency.”}}
\end{itemize}

In these examples, the interviewer understood the participant's statement and feelings and responded with a short but accurate response, making the conversation feel more natural and engaging. Evaluation of all 41 interview transcripts shows that no semantic mismatches or tone violations were found, confirming the system's reliability.

\section{Performance Evaluation}

Three key performance indicators of AI-powered interviewer were evaluated to study the effectiveness. The key performance indicators are 

\begin{itemize}
  \item Cognitive and Emotional Engagement: How well does the system keep the user's attention and engaged?
  
  \item Question Relevance and Coherence: Are the questions logical and relevant to the conversation?

  \item Overall User Satisfaction and Comparative Experience: How satisfied are users with the experience, and how does it compare to traditional human-led interviews?
\end{itemize}

These performance indicators show how well the system can have a meaningful conversation, ask good questions, and provide good user experience.

Together, these indicators provide a robust framework for evaluating the usability, adaptability, and success of the proposed AI-interviewer in conducting semi-structured qualitative interviews.

As presented in Table \ref{tab:descriptiveStats}, descriptive statistics show consistently high mean scores across all three indicators, indicating a strong positive experience among participants. Low standard deviations suggest a high level of agreement, and the skewness and kurtosis values fall within acceptable ranges, supporting the use of parametric analysis.

\begin{table}[htbp]
\centering
\caption{Descriptive statistics of performance evaluation factors}
\resizebox{\textwidth}{!}{%
\begin{tabular}{|l|c|c|c|c|c|c|c|c|c|c|}
\hline
\textbf{Variable} & \textbf{N} & \textbf{Min} & \textbf{Max} & \textbf{Mean} & \textbf{Std. Dev.} & \textbf{Variance} & \textbf{Skewness} & \textbf{Skew. SE} & \textbf{Kurtosis} & \textbf{Kurt. SE} \\
\hline
\makecell[l]{Question Relevance \\and Coherence} & 41 & 3.67 & 5.00 & 4.61 & 0.4825 & 0.23 & -0.519 & 0.369 & -1.615 & 0.724 \\
\hline
\makecell[l]{Cognitive and \\Emotional Engagement} & 41 & 3.33 & 5.00 & 4.33 & 0.3496 & 0.12 & -0.410 & 0.369 & 0.986 & 0.724 \\
\hline
\makecell[l]{Overall User Satisfaction and\\Comparative Experience} & 41 & 4.00 & 5.00 & 4.45 & 0.2917 & 0.09 & 0.003 & 0.369 & 0.080 & 0.724 \\
\hline
\end{tabular}%
}
\label{tab:descriptiveStats}
\end{table}

The regression model summary in Table \ref{tab:ModelSummary} indicates a moderate relationship between predictors and user satisfaction, with R = 0.499 and R² = 0.249. This means that approximately 24.9\% of the variation in satisfaction is explained by engagement and question relevance. Although not a strong predictive model, it is acceptable in exploratory and human-computer interaction studies, where user satisfaction is shaped by multiple contextual variables.

The statistical significance of the model (F = 6.289, p = 0.004) is confirmed by the ANOVA results in Table \ref{tab:Anova}, which indicates that cognitive engagement and question relevance together meaningfully contribute to predicting satisfaction.

\begin{table}[htbp]
\centering
\caption{Regression Model Summary Showing the Strength of the Relationship Between Predictors and User Satisfaction}
\label{tab:ModelSummary}
\begin{tabular}{|c|c|c|c|c|}
\hline
\textbf{Model} & \textbf{R} & \textbf{R Square} & \textbf{Adjusted R Square} & \textbf{Std. Error of the Estimate} \\
\hline
1 & .499\textsuperscript{a} & .249 & .209 & .2594 \\
\hline
\end{tabular}

\begin{flushleft}
\textsuperscript{a} Predictors: (Constant), Cognitive and Emotional Engagement, Question Relevance and Coherence
\end{flushleft}
\end{table}

\begin{table}[htbp]
\centering
\caption{ANOVA Table Indicating the Overall Significance of the Regression Model}
\label{tab:Anova}
\begin{tabular}{|c|l|c|c|c|c|c|}
\hline
\textbf{Model} & \textbf{} & \textbf{Sum of Squares} & \textbf{df} & \textbf{Mean Square} & \textbf{F} & \textbf{Sig.} \\
\hline
\multirow{3}{*}{1} 
& Regression & .846 & 2 & .423 & 6.289 & .004\textsuperscript{a} \\
& Residual   & 2.556 & 38 & .067 &       &       \\
& Total      & 3.402 & 40 &       &       &       \\
\hline
\end{tabular}

\begin{flushleft}
\textsuperscript{} ANOVA Dependent Variable: Overall User Satisfaction and Comparative Experience \\
\textsuperscript{a} Predictors: (Constant), Cognitive and Emotional Engagement, Question Relevance and Coherence
\end{flushleft}
\end{table}

\begin{table}[htbp]
\centering
\caption{Regression Coefficients Showing the Individual Contribution of Predictors to User Satisfaction}
\label{tab:correlation}
\begin{tabular}{|c|l|c|c|c|c|c|}
\hline
\textbf{Model} & & \multicolumn{2}{c|}{\makecell{\textbf{Unstandardized}\\\textbf{Coefficients}}} & \makecell{\textbf{Standardized}\\\textbf{Coefficients}} & \textbf{t} & \textbf{Sig.} \\

\cline{3-4} \cline{5-5}
 & & \textbf{B} & \textbf{Std. Error} & \textbf{Beta} & & \\
\hline
\multirow{3}{*}{1} 
& (Constant) & 2.965 & 0.676 &       & 4.384 & $<$.001 \\
& \makecell[l]{Question Relevance and\\Coherence} & -0.055 & 0.085 & -0.091 & -0.645 & 0.523 \\
& \makecell[l]{Cognitive and Emotional\\Engagement} & 0.402 & 0.118 & 0.481 & 3.407 & 0.002 \\
\hline
\end{tabular}

\begin{flushleft}
\textsuperscript{} Dependent Variable: Overall User Satisfaction and Comparative Experience
\end{flushleft}
\end{table}

More detailed analysis of the regression coefficients in Table \ref{tab:correlation} indicates that Cognitive and Emotional Engagement is a significant positive predictor (B~=~0.402, p~=~0.002), emphasizing why it is important to keep users emotionally and mentally during the interview conversation. Whereas, Question Relevance and Coherence, although rated highly in descriptive terms failed to significantly predict satisfaction when analyzed independently (B~=~-0.055, p~=~0.523), suggesting that once users are engaged, question relevance may be viewed as a baseline expectation rather than a standout factor. Generally, the performance evaluation indicates that the AI-powered interviewer worked well. It kept users engaged and satisfied, adapting to their responses, making conversation personal and meaningful. These results show that the interviewer can be an effective tool for the collection of qualitative data via interviews.

\section{Analysis and Discussion}

Many participants from diverse backgrounds were highly satisfied with the interview experience. The participants found the question to be relevant and clear; they generally enjoyed how the conversation flowed. The interviewer was rated high for its adaptability as it continued the conversation based on prior conversations, which made the conversation feel natural and easy to follow. Some participants also loved the interviewer's neutral and non-judgmental tone, because it made them feel comfortable discussing sensitive topics. Participants were also happy with the actual questions, which they rated as being clear and easy to understand. Generally, the study indicates that the AI-powered interviewer is a promising alternative way to conduct qualitative interviews.

However, differences in preferences were observed among the participants. Some participants preferred traditional interviews, while others liked the AI-powered interviews. Older participants were found to prefer traditional interviews because they provide personal connection and felt more natural. Contrariwise, younger participants were more open to AI-powered interviews. Additionally, introverted participants preferred AI-powered interviews over traditional interviews because they felt more comfortable and less pressured. Several participants thought the AI-powered interviews were good for discussing sensitive topics like racism and discrimination because they are neutral and non-judgmental. 

The AI-powered interviewer's response messages were appreciated because they were brief and relevant to what people said. They also pointed out that this made the conversation feel natural and easy to follow. Also, they added that the questions were easy to understand and kept people interested throughout the session. This indicates that the interviewer was designed efficiently to conduct qualitative interviews. Even though the initial pilot study of the interviewer showed promising results, it is important to acknowledge some limitations in the current version. The current version of the interviewer depends heavily on well-crafted prompts that could fail due to prompt weaknesses when the LLM model is swapped. Also, the smaller LLaMA model used has context memory limitations, so if the interview lasts for several hours, the interviewer may not remember the entire conversation.

\section{Conclusion}

This study presented a modular AI-powered interviewer to help conduct qualitative interviews. The interviewer was designed to conduct adaptive, secure, and context-aware qualitative interview sessions. The interviewer was implemented using a local LLM (LLaMA 3.2, 3B). The pilot test results showed that participants were very satisfied with the interview (average score: 4.52), engaged and focused (average score: 4.39), and thought the interview questions were very relevant (average score: 4.41). The study found that engagement was a strong predictor of satisfaction, explaining nearly 25\% of the difference in satisfaction scores.

These findings suggest that the interviewer has great potential for conducting scalable and privacy-focused qualitative data collection. The modular design makes it easy to modify and use across different languages, cultures, and topics. In the future, it would be useful to integrate an automated tool to analyze and summarize participant responses in real-time for immediate insight, as proposed in \cite{adeseye2025} and \cite{adeseye2025ichms}, while keeping the data private. Expanding context memory and enhancing multilingual capabilities are important areas to consider for future development as well.

\end{document}